\documentclass[fleqn,usenatbib]{mnras}

\usepackage{newtxtext,newtxmath}

\usepackage[T1]{fontenc}
\usepackage{pdflscape}

\DeclareRobustCommand{\VAN}[3]{#2}
\let\VANthebibliography\thebibliography
\def\thebibliography{\DeclareRobustCommand{\VAN}[3]{##3}\VANthebibliography}

\usepackage{xcolor}

\definecolor{brightpink}{rgb}{1.0, 0.0, 0.5}

\usepackage{graphicx}	
\usepackage{amsmath}	

\title[GRB Photo-z Estimation]{Photometric Redshift Estimation for Gamma-Ray Bursts from the Early Universe}

\author[H. M. Fausey et al.]
{H.M. Fausey,$^{1}$\thanks{E-mail: hfausey@gwu.edu}
A.J. van der Horst,$^{1}$
N.E. White,$^{1}$
M. Seiffert,$^{2}$
P. Willems,$^{2}$
E.T. Young,$^{3}$
D.A. Kann,$^{4}$\thanks{deceased}\newauthor
G. Ghirlanda,$^{5,6}$
R. Salvaterra,$^{7}$
N.R. Tanvir,$^{8}$
A. Levan,$^{9}$
M. Moss,$^{1}$
T-C. Chang,$^{2}$
A. Fruchter,$^{10}$
S. Guiriec,$^{1}$\newauthor
D. H. Hartmann,$^{11}$
C. Kouveliotou,$^{1}$
J. Granot,$^{12,13,1}$
A. Lidz$^{14}$
\\
$^{1}$Department of Physics, George Washington University, 725 21st Street NW, Washington, DC 20052, USA\\
$^{2}$Jet Propulsion Laboratory, California Institute of Technology, 4800 Oak Grove Dr, Pasadena, CA 91109, USA\\
$^3$USRA, 425 3rd Street SW, Suite 950, Washington, DC 20024, USA\\
$^{4}$Instituto de Astrofisica de Andalucia, Glorieta de la Astronomia, E-18008 Granada, Spain\\
$^{5}$INAF, Osservatorio Astronomico di Brera, Via E Bianchi 46, 23807 Merate (LC), Italy\\
$^{6}$INFN -- Sezione di Milano-Bicocca, piazza della Scienza 3, I-20126 Milano (MI), Italy\\
$^{7}$INAF IASF-Milano, Via Alfonso Corti 12, I-20133 Milano, Italy\\
$^{8}$School of Physics and Astronomy, University of Leicester, University Rd, Leicester, LE1 7RH, UK\\
$^{9}$Department of Astrophysics/IMAPP, Radboud University, PO Box 9010, 6500 GL, The Netherlands\\
$^{10}$Space Telescope Science Institute, 3700 San Martin Dr, Baltimore, MD 21218, USA\\
$^{11}$Department of Physics \& Astronomy, Clemson University, Kinard Lab of Physics, Clemson, SC 29634, USA\\
$^{12}$Astrophysics Research Center of the Open University (ARCO), The Open University of Israel, P.O Box 808, Ra’anana 4353701, Israel\\
$^{13}$Department of Natural Sciences, The Open University of Israel, P.O Box 808, Ra’anana 4353701, Israel\\
$^{14}$Department of Physics \& Astronomy, University of Pennsylvania, 209 S. 33rd Street, Philadelphia, PA 19104, USA
}

\date{Accepted XXX. Received YYY; in original form ZZZ}

\pubyear{2023}

\begin{document}
\label{firstpage}
\pagerange{\pageref{firstpage}--\pageref{lastpage}}
\maketitle

\begin{abstract}
Future detection of high-redshift gamma-ray bursts (GRBs) will be an important tool for studying the early Universe.
Fast and accurate redshift estimation for detected GRBs is key for encouraging rapid follow-up observations by ground- and space-based telescopes.
Low-redshift dusty interlopers pose the biggest challenge for GRB redshift estimation using broad photometric bands, as their high extinction can mimic a high-redshift GRB.
To assess false alarms of high-redshift GRB photometric measurements, we simulate and fit a variety of GRBs using \texttt{phozzy}, a simulation code developed to estimate GRB photometric redshifts, and test the ability to distinguish between high- and low-redshift GRBs when using simultaneously observed photometric bands.
We run the code with the wavelength bands and instrument parameters for the Photo-z Infrared Telescope (PIRT), an instrument designed for the \emph{Gamow} mission concept. We explore various distributions of host galaxy extinction as a function of redshift, and their effect on the completeness and purity of a high-redshift GRB search with the PIRT. 
We find that for assumptions based on current observations, the completeness and purity range from $\sim 82$ to $88\%$ and from $\sim 84$ to $>99\%$, respectively. For the priors optimized to reduce false positives, only $\sim 0.6\%$ of low-redshift GRBs will be mistaken as a high-redshift one, corresponding to $\sim 1$ false alarm per 500 detected GRBs.
\end{abstract}




\section{Introduction}

Gamma-ray bursts (GRBs) are the most electromagnetically luminous events in the Universe.
They are divided into two classes \citep{Mazets1981, Kouveliotou1993}: short GRBs are thought to arise from compact object mergers \citep{Eichler1989,Narayan1992}, while long GRBs result from the core collapse of massive stars \citep{Woosley1993}.
While both types of GRB are exceedingly bright, long GRBs are the most luminous \citep{Hjorth2003}, with the brightest having isotropic-equivalent luminosities in excess of $ 10^{54}$ ergs/s \citep{Frederiks2013}, allowing them to potentially be detected out to redshifts as high as $\sim 20$ \citep{Lamb2000}.
Given their simple power-law spectra and extreme luminosities, long GRBs are ideal probes of the high-redshift Universe.
They can be used to trace the chemical evolution of the Universe \citep{Savaglio2006,Thone2013,Sparre2014, Saccardi2023}, study early star formation in the initial mass function \citep{Lloyd2002, Fryer2022}, Population III stars \citep{Lloyd2002,Campisi2011}, and constrain the Epoch of Reionization by examining hydrogen in the high-redshift intergalactic medium \citep[IGM;][]{MiraldaEscude1998, Totani2006, Hartoog2015, Lidz2021}.
For recent reviews of GRBs, see \citet{Salvaterra2015, Levan2016, Schady2017, vanEerten2018, Luongo2021}.

The Epoch of Reionization is theorized to have ended around redshift $z \sim 6$, or 1 billion years after the Big Bang \citep{Totani2006}. GRBs at redshifts higher than $z \sim 6$ will facilitate the study of star formation and chemical evolution at earlier stages in the evolution of the Universe.
Fewer than 10 GRBs with $z > 6$ have been found to date \citep{Tanvir2009, Cucchiara2011, Salvaterra2015, Tanvir2018}. 
While there are multiple missions dedicated to GRB detection and science, such as the \emph{Neil Gehrels Swift Observatory} \citep[\emph{Swift};][]{Gehrels2004}, the \emph{Fermi Gamma-ray Space Telescope} \citep[\emph{Fermi};][]{Atwood2009}, and KONUS-\emph{Wind} \citep{Aptekar1995}, these missions are not optimized for detecting high-redshift GRBs.
The community has clearly recognized this absence, as a legion of high-z GRB missions have been proposed over more than a decade: \emph{Xenia} \citep{Kouveliotou2008}, \emph{Joint Astrophysics Nascent Universe Survey} \citep[\emph{JANUS};][]{Burrows2010}, \emph{Energetic X-ray Imaging Survey Telescope} \citep[\emph{EXIST};][]{Grindlay2010}, \emph{Origin} \citep{Piro2011}, \emph{High-z gamma-ray bursts for unraveling the dark ages mission} \citep[\emph{HiZ-GUNDAM};][]{Yonetoku2014}, \emph{Gamow Explorer} \citep[\emph{Gamow};][]{White2021}, \emph{Transient High-Energy Sky and Early Universe Surveyor} \citep[\emph{THESEUS};][]{Amati2021}, and \emph{Space Variable Objects Monitor} \citep[\emph{SVOM};][]{Atteia2022}. 
So far, only \emph{SVOM} has been selected to become a mission, but it will be crucial to launch missions and instruments designed to detect high-z GRBs if we are to fully study the environments and evolution of the early Universe using GRBs as probes.

A major component of any future high-z GRB mission will be to alert the community for rapid follow-up observations of high-z GRBs.
Fast and accurate on-board redshift estimation will be vital to their success.
There are two ways to estimate redshift, using photometry and spectroscopy. 
Photometric redshift estimation relies on Lyman-$\alpha$ (Ly$\alpha$) blanketing, a sharp loss of flux due to a multitude of absorption lines from neutral hydrogen throughout the IGM \citep{Madau1996, Madau1995}.
Spectroscopic redshift estimates are more accurate but require a significantly longer observing time, as they rely on measuring the flux through many small wavelength bins. Photometry uses broad bands, so it is faster than spectroscopy but less accurate for redshift estimation.
Because of its speed and technical constraints, photometry is more practical for rapidly alerting the community to potential high-z GRBs.
However, photometry is susceptible to misidentifying a high-extinction GRB as a high-redshift one \citep{Curran2008}.
Extinction can cause a seemingly steep loss in flux when using broad photometric bands, as it has a larger effect at shorter, bluer wavelengths \citep{Fitzpatrick1989, Cardelli1989, Pei1992, Schady2012}.
For high-extinction GRBs where this effect is more pronounced, it can be difficult to distinguish it from a high-z GRB.
Finding as large a number of true high-z GRBs as possible (completeness), while minimizing the number of low-z high-extinction GRBs causing false alarms (purity), will establish the potential success of a high-z GRB mission.

Here we present an in-depth examination into our ability to correctly identify high-z GRBs using \texttt{phozzy}, a photometric redshift estimation code capable of simulating and fitting photometric measurements, and estimating instrument performance for future high-z GRB missions.
The code can be applied to any instrument that makes use of any number of simultaneously observed photometric bands for redshift estimation.
As an example, we use the channels proposed for the Photo-z Infrared Telescope \citep[PIRT;][]{Seiffert2021, White2021} onboard \emph{Gamow}, a high-z GRB mission proposed to the 2021 NASA MIDEX call \citep{White2021}.
The PIRT is an instrument designed to identify high-z GRBs within 100 seconds and send an alert within 1000 seconds of a GRB trigger \citep{White2021}.
We use the PIRT to estimate how accurately we can differentiate between true high-z GRBs, and low-z high-extinction GRBs.

In Section \ref{sec:methodology}, we will discuss details about the code, including the models, simulated GRB generation, and fitting used for the simulations.
We present the results of the simulations in Section \ref{sec:results}, including the key metrics used for estimating instrument performance. In Sections \ref{sec:discussion} and \ref{sec:conclusion}, we consider the implications of the results for future GRB missions and summarize our findings.

\section{Methodology and Modelling Code}
\label{sec:methodology}

In this section, we discuss the motivation and details behind our spectral model and the fitting method. 
We then explain the choices of distributions for both the input parameters of the simulated GRBs and priors for the fitting.
Finally, we describe the structure of the code and how it functions.

\subsection{Model}
\label{sec:model}
We assume that the optical to near-infrared (NIR) regime of a GRB afterglow spectrum can be modelled by a single power-law function of flux versus frequency or wavelength \citep{Sari1998}.
Two major effects are applied to the spectrum: host galaxy extinction and intergalactic attenuation.

Host galaxy extinction is the absorption and scattering of light due to gas and dust along the line of sight in the GRB host galaxy \citep{Pei1992, Klose2000,Zafar2011,Greiner2011, Covino2013, Bolmer2018}.
Extinction is most prominent in the ultra-violet (UV) and optical regimes, and primarily affects shorter, bluer wavelengths \citep{Fitzpatrick1989, Cardelli1989, Pei1992}. For GRB host galaxies, there are typically three templates used: the Small Magellanic Cloud (SMC), Large Magellanic Cloud (LMC) and Milky-Way extinction models developed by \cite{Pei1992}.
Our code includes all three models, but we chose to use the SMC model for the simulations presented in this paper because it is most consistent with observations for GRB host galaxies, which tend not to have a bump at 2175~\AA~\citep[e.g.,][]{Schady2012}. The 2175~\AA~ bump was originally attributed to graphite and silicate grains \citep{Mathis1977, Drain1984, Pei1992, Schady2012}, or a mix of carbonaceous grains with polycyclic aromatic hydrocarbons (PAHs) \citep{Li2001, Weingartner2001, Draine2007, Fischera2008}. However, recent work suggest that PAHs alone may be the cause \citep{Shivaei2022, Hensley2023, Lin2023}.
 
Intergalactic attenuation is applied to the extincted spectrum using a model developed by \cite{Meiksin2006}, which is based on work by \cite{Madau1995}. 
Intergalactic attenuation is the photoelectric absorption and resonant scattering by hydrogen gas in the intergalactic medium \citep{Madau1995, Meiksin2006}.
It is dominated by the Lyman-$\alpha$ (Ly$\alpha$) line, which has a wavelength of $\lambda_\alpha = 1216$~\AA, corresponding to the transition between the first and second energy levels of the hydrogen atom \citep{Madau1995, Madau1996, Meiksin2006}.
This results in a Ly$\alpha$ forest -- a region of spectra `blanketed' by a multitude of Ly$\alpha$ (and other higher order Lyman series) absorption lines due to intervening hydrogen \citep{Madau1995,Madau1996}.
The red edge of the Ly$\alpha$ forest features a steep drop-off in spectra that occurs at a wavelength of $\lambda_\alpha(1+z)$, so it is heavily dependent on the redshift \citep{Madau1995,Madau1996}.
This Ly$\alpha$ drop-off is a key marker for estimating the redshift of extra-galactic objects \citep{Steidel1992,Steidel1996,Madau1996,Kruhler2011}, and is the main feature used by \texttt{phozzy} to determine the photometric redshift of a GRB.

Overall the model is controlled by 4 parameters: the flux normalization $A$, the spectral index $\beta$, the redshift $z$, and the extinction $E_{B-V}$. The afterglow spectral flux is given by $A$ and $\beta$, while $E_{B-V}$ and $z$ influence the spectral shape through effects related to extinction and redshift, respectively.

\subsection{Fitting with MCMC}
\label{sec:mcmc}

The code runs a Markov Chain Monte Carlo (MCMC) fitting method using \texttt{emcee}, a python package developed by \cite{emcee}.
The MCMC method is a stochastic process that estimates parameters using posterior distributions \citep{MCMCtextbook}.
The posteriors are determined by a likelihood function based on the fit to the data and priors, probability distributions for the parameters based on a priori information \citep{MCMCtextbook}.

The \texttt{phozzy} package uses a Bayesian likelihood function 
$$\log(\mathcal{L}) = -\frac{\chi^2}{2}$$
where $\mathcal{L}$ is the likelihood, and $\chi^2$ is defined as
$$\chi^2 = \sum_1^n \frac{(m_i - f_i)^2}{\sigma_i^2}$$
where $m_i$ and $f_i$ are the measured and fit average fluxes across each photometric band $i$, respectively, and $\sigma_i$ are the uncertainties of the measurements. 
The priors used in the Bayesian analysis are outlined in Section \ref{sec:priors}.
In the photo-z code, a set of 50 walkers are given a 250-step `burn-in' phase allowing them to settle into a preferred, stable region of parameter space before starting its 500-step `production' phase from which the posterior distributions are created \citep{MCMCtextbook}.
Once the runs are complete, the code takes the final positions of all 50 walkers for further analysis.

Even though there are only 4 parameters, the parameter space is complex due to effects from host galaxy extinction and intergalactic attenuation. Furthermore, the number of photometric bands is typically not much larger than the number of free parameters. 
This makes fitting prone to getting stuck in local minima instead of finding a global minimum.
An MCMC fitting method is helpful for complex parameter spaces because it is stochastic and less likely to get stuck in a local minimum.
With an MCMC method we can take the posterior distributions and the final positions of the walkers to identify multiple minima in parameter space if they exist.

\subsection{Parameter Priors and Inputs}
\label{sec:i&p}
MCMC fitting methods allow for the input of a set of priors, or probability distributions for the parameters that can inform the posteriors.
These simulations also require distributions for the input parameters that will be used to simulate the photometric band fluxes.
Here we describe the parameter distributions used for both priors and input.

\subsubsection{Priors}
\label{sec:priors}
For the flux normalization $A$ and redshift $z$, the code includes a simple positive uniform prior.
We considered including a redshift prior based on an expected redshift distribution of GRBs the would have been observed with Gamow \citep[based on][]{Ghirlanda2015,Ghirlanda2021,Ghirlanda2022}, but found that it resulted in a large portion of high-z GRBs being confused as low-z GRBs. 
We instead use this expected distribution for the GRB input redshifts (see Section \ref{sec:inputs})
The spectral index is given a Gaussian prior centered on $\beta = 0.7$ with a standard deviation of $0.2$, which is based on a large sample of observed optical afterglow spectral indices \citep{Li2015}.

The code includes multiple priors for the host galaxy extinction $E_{B-V}$, which are explored in the simulations presented in this paper:
\begin{itemize}
    \item `none' -- extinction is assumed to be 0 and is not fit as a free parameter;
    \item `basic' -- a single exponential prior based on results from \cite{Covino2013};
    \item `evolving' -- a more complex exponential prior that combines results from \cite{Covino2013}, \cite{Bolmer2018} and \cite{Greiner2011}, in which the exponential constant changes depending on the redshift.
\end{itemize}

In their study of \emph{Swift} GRBs, \citet{Covino2013} found that most GRBs have a relatively low host galaxy extinction, with a quickly diminishing number of GRBs at higher extinctions. We model our `basic' prior off of this \citep{Covino2013} data by fitting their extinction distribution with an exponential function.

A later study performed by \citet{Bolmer2018} found that higher redshift GRBs have significantly lower extinction compared to those of lower-redshift GRBs, with GRBs with $2 < z < 4$ all having $A_{\rm V} < 3$, and GRBs with $z > 4$ having $A_{\rm V} < 0.5$. 
The same $A_{\rm V}$ thresholds appear when combining the GRBs examined by \citet{Bolmer2018} and \citet{Covino2013} with an additional GRB host galaxy extinction study done by \citet{Greiner2011}.
To account for this relationship between redshift and extinction, we also created an `evolving' extinction prior for which we fit the extinction distributions for the $z < 2$, $2 < z < 4$, and $z > 4$ data sets with an exponential function, and use the different exponents for their corresponding redshift ranges (See Table \ref{tab:EbvPriors}).

We also include the option for 2 different sets of upper limits on the `evolving' extinction prior, based on the $A_{\rm V}$ thresholds for different redshift ranges in the \citet{Bolmer2018}, \citet{Covino2013}, and \citet{Greiner2011} GRB data (See Table \ref{tab:EbvPriors}).

\begin{table*}
    \centering
    \caption{Table of exponents for each exponential prior of the host galaxy extinction, including the various exponents for different redshift ranges in the `evolving' prior. Note that the `evolving' prior exponents get steeper in higher redshifts ranges due to the increasingly quick drop-off in $A_{\rm V}$ at higher redshifts. The table also shows the different $A_{\rm V}$ and $E_{B-V}$ upper limits included in the code, based on data from \citet{Bolmer2018}, \citet{Covino2013}, and \citet{Greiner2011}.}
    \begin{tabular}{c|c|c|c|c}
    \hline
        Prior type & $z$ & $E_{B-V}$ exponent & upper limit 1 & upper limit 2\\
         & & & 
         \begin{tabular}{c|c}
         $A_{\rm V}$ & $E_{B-V}$     
         \end{tabular} 
         &
         \begin{tabular}{c|c}
         $A_{\rm V}$ & $E_{B-V}$     
         \end{tabular}\\
        \hline
        basic & $\geq 0$ & 4.28 & \begin{tabular}{c|c}
         --- & ---
         \end{tabular}
         &
         \begin{tabular}{c|c}
         --- & ---
         \end{tabular}\\
         evolving & 0 - 2 & 6.9 &
         \begin{tabular}{c|c}
         6 & 2.05
         \end{tabular}
         &
         \begin{tabular}{c|c}
         6 & 2.05
         \end{tabular}\\ 
         evolving & 2 - 4 & 12.6 & \begin{tabular}{c|c}
         3 & 1.02
         \end{tabular}
         &
         \begin{tabular}{c|c}
         3 & 1.02
         \end{tabular}\\
         evolving &  > 4 & 36.2 & \begin{tabular}{c|c}
         1 & 0.34
         \end{tabular}
         &
         \begin{tabular}{c|c}
         0.5 & 0.17
         \end{tabular}\\
         \hline
    \end{tabular}
    \label{tab:EbvPriors}
\end{table*}

\subsubsection{Inputs}
\label{sec:inputs}
When selecting the input parameters for each simulated GRB, the code can pull from any of the priors for each respective parameter, but with two additions: a log-normal redshift distribution and a Gaussian flux distribution with a mean flux determined by the input redshift.

We include a log-normal input redshift distribution ($\mu = 0.8$, $\sigma = 0.55$) based on an expected redshift distribution for GRBs observed by Gamow and based on previous work found in \citet{Ghirlanda2022,Ghirlanda2021,Ghirlanda2015}. 
We chose not to include this distribution as a prior because it negatively impacted the fitting methods' ability to identify high-z GRBs because the distribution is concentrated around low redshifts, and increases the likelihood that it would identify high-z GRBs as low-z GRBs.
However, as an input distribution it can estimate how accurate the photo-z measurements will be for a general population of GRBs.
False positives due to low-z dusty interlopers are a great concern for a high-z GRB mission, so it is vital to ensure that low-redshift GRBs are correctly categorized.
This is especially important since the vast majority of GRBs will likely have $z < 5$, so even a small percentage of false positives could result in more false positives than true high-z detections.

 For the input flux we use a Gaussian prior for the log of the flux in $\mu$Jy ($\mu = 6.18, \sigma = 2.65$) based on the expected brightness distribution for GRBs at redshift 10 (Kann et al., in preparation). 
 The flux is then adjusted for the redshift of the GRB by using the ratio of luminosity distances \citep{Weinberg1972} to determine what the brightness would be for a similar GRB at a different redshift.
 Note that we do not account for the lower intrinsic luminosities for lower-redshift GRBs \citep{Petrosian2015, Pescalli2016, Lloyd2019, Banerjee2022}.
 This flux distribution may also be optimistic for \emph{Gamow}, since it may detect more faint GRBs with dimmer afterglows (Kann et al., in preparation). If this is the case, this distribution may underestimate the rate at which the PIRT will observe high-redshift GRBs with low fluxes, which tend to have less certain photometric redshift estimations.
However, we do not use this distribution as a prior, so it has a limited impact on the fitting itself. 
As an input distribution, we chose to rely on the added uncertainty in the band measurements to envelop any discrepancies.

\subsection{Data \& Code Structure}

The code first generates parameters for the desired number of GRBs.
These sets of parameters are randomly generated by pulling from the input distributions specified by the user (see Section \ref{sec:inputs}), to create a set of simulated GRB spectra using the model laid out in Section \ref{sec:model}.
From these spectra, the flux measurements for each of the specified photometric bands are determined by finding the average integrated flux across the entire band, and then perturbed according to the assumed statistical uncertainty added in quadrature to the estimated instrumental noise.
If the measured flux is below the given detection limit for all photometric bands, the corresponding set of parameters, GRB spectrum, and band measurements are recreated to ensure that all simulated GRBs would be considered detections.

\begin{figure}
    \centering
    \includegraphics[width=\linewidth]{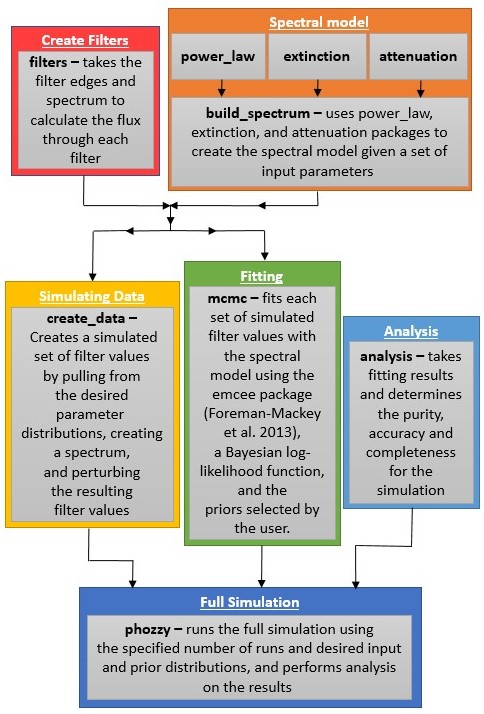}
    \caption{The structure of \texttt{phozzy}. Diagram shows the different modules and in which other modules of the code they are used. The main module \texttt{phozzy} uses the \texttt{create\_data}, \texttt{mcmc}, and \texttt{analysis} modules to first create a number of simulated GRB photometric band measurements, then perform fits on each set, and finally, compile all results and determine the completeness, purity, and accuracy metrics for the results.}
    \label{fig:codeStructure}
\end{figure}

The perturbed fluxes are fit using the MCMC fitting method described in Section \ref{sec:mcmc} using the priors selected by the user (see Section \ref{sec:priors}).
The final fits, posteriors, and positions of all walkers in parameter space for each of the fit are saved for further analysis. The full code structure is given in Figure~\ref{fig:codeStructure}.

For the example simulations, we input the 5 simultaneously observed optical-NIR bands proposed for the PIRT, which have wavelength ranges of $0.50-0.64$, $0.64 - 0.87$, $0.87 - 1.2$, $1.2 - 1.7$, and $1.7 - 2.4$ $\mu m$. 
These bands were part of the final design reported in the \emph{Gamow Explorer}'s NASA MIDEX proposal, and are updated versions of the ranges from \citet{Seiffert2021} and \citet{White2021}. 
These band edges correspond to the wavelengths of the Ly$\alpha$ breaks for redshifts of 3.1, 4.3, 6.2, 8.9, 13.0, and 18.7, respectively.

\texttt{phozzy} allows for a large amount of customization. 
It includes inputs for the instrument parameters, such as band edges, statistical uncertainties, the 1$\sigma$ instrument noise value, and an $n$$\sigma$ detection limit, where $n$ is how many multiples above the 1$\sigma$ instrument noise level a measurement must be to be considered a detection in that band. 
It also allows the user to indicate the desired host galaxy extinction model, and the parameter input and prior distributions.
The code can be used for instruments with bands that are not observed simultaneously, but the user must account for the additional effects and uncertainty that arise when interpolating flux measurements (see Section \ref{sec:discussion}).
For each of the PIRT simulations we create and fit sets of 500 GRBs, and assume a statistical uncertainty of 5\% and 3~$\mu$Jy of 1$\sigma$ instrument noise. 
The results for the PIRT simulations are presented in Section \ref{sec:results}.

\section{Results}
\label{sec:results}

Here we present the results of the \texttt{phozzy} simulations for the \emph{Gamow} PIRT instrument, using 3 metrics to assess the instrument performance for categorizing high- vs. low-redshift GRBs, and retrieving the redshift:

\begin{itemize}
    \item completeness -- for GRBs above redshift 5, how many are correctly identified as high-redshift ($z>5$) GRBs;
    \item purity -- for GRBs below redshift 5, how many are correctly identified as low-redshift ($z<5$) GRBs;
    \item accuracy -- for GRBs above redshift 5, how often does the fitting method return a redshift within 10\% and 20\% of the input redshift.
\end{itemize}

\begin{figure}
    \centering
    \includegraphics[width = \linewidth]{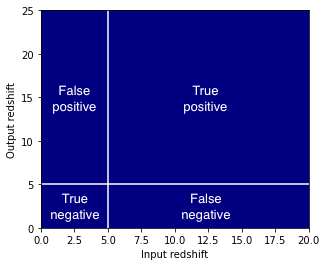}
    \caption{Regions defined as true positive, false positive, true negative, and false negative based on the input and output redshift results. For these simulations, we consider $z=5$ the threshold between low and high redshift GRBs (white lines).}
    \label{fig:grid}
\end{figure}

The accuracy metric determines how well the fits retrieve the true redshift, while completeness determines the likelihood of a true positive (as opposed to a false negative) for high-redshift GRBs, and purity gives the likelihood of a true negative (as opposed to a false positive) for low-redshift GRBs (see Figure~\ref{fig:grid}). 
We estimate these metrics by running fits on 500 randomly generated GRB spectra with a variety of input distributions and priors. 
For these simulations we define high-redshift as $z > 5$, but the code allows this threshold to be changed by the user.

For these simulations we draw from two different redshift distributions for varying purposes: a uniform redshift distribution for which all redshifts between 0 and 20 are equally likely, and an expected redshift distribution for Gamow that was created using previous work found in \citet{Ghirlanda2022,Ghirlanda2021,Ghirlanda2015}.

The uniform redshift distribution is mainly used for measuring completeness, while the measured redshift distribution is mainly used for measuring purity. 
To estimate completeness, we need to establish the instrument performance at all redshifts. The measured redshift distribution has very few high-z GRBs, and those that do occur are concentrated towards the lower end between $z = 5-6$, so it supplies very little information about the instrument performance if, for example, a $z=10$ GRB were to be observed. The uniform distribution is evenly distributed across all redshifts, so it truly tests the instrument performance for all redshifts. 
For estimating purity, we do need to take the measured redshift distribution into account. In this case, we need to know how many misidentified low-z GRBs there are, which implies using a distribution similar to what we expect to observe. 
For estimating accuracy, we use and present the results for both distributions.

For both of these redshift distributions we run a variety of configurations for input and fitting prior extinction distributions:
\begin{itemize}
    \item Input: no extinction, Fitting: no extinction -- acts as a baseline and makes sure the code is working properly; from here on, this will be denoted as the `no extinction baseline'.
    \item Input: no extinction, Fitting: basic extinction prior -- acts as a baseline for how well the instrument would do in the best case scenario where there is no host galaxy extinction for any GRBs; this shows how adding extinction as a free parameter affects the ability to retrieve the redshift, and will be denoted as the `extinction prior baseline'.
    \item Input: basic extinction, Fitting: basic extinction prior -- shows how well the instrument performs for an expected distribution of GRBs when using the basic fitting prior.
    \item Input: `evolving' extinction, Fitting: `evolving' extinction prior -- shows how well the instrument performs for an expected distribution of GRBs when using the `evolving' fitting prior.
    \item Input: `evolving' extinction with upper limit 1, Fitting; `evolving' extinction with upper limit 1 -- shows how results change when adding a lightly constrained upper limit on host galaxy extinction when using the `evolving' extinction prior.
    \item Input: `evolving' extinction with upper limit 2, Fitting: `evolving' extinction with upper limit 2 --  shows how results change when adding a more constrained upper limit on host galaxy extinction when using the `evolving' extinction prior. 
\end{itemize}

While the runs with no extinction for the inputs and/or priors form a baseline, the others determine which set of priors will be best for correctly identifying low- and high-redshift GRBs for the \emph{Gamow} PIRT instrument specifications (observing bands, instrument noise, etc.).
The results for all runs are summarized in Table~\ref{tab:allResults}, and displayed using density plots showing the input versus retrieved redshift for all runs in Appendix A. The results and implications for the completeness, purity, and accuracy are detailed in the following subsections.

\begin{table*}
\caption{Results from all simulations, listing redshift (z) and host galaxy extinction ($E_{B-V}$) inputs and priors, as well as the completeness (for GRBs above redshift 5, how many are correctly identified as high-z GRBs) and purity (for GRBs below redshift 5, how many are correctly identified as low-z GRBs). Baseline runs are italicized, and the most relevant statistic in terms of completeness and purity are in bold.}
    \begin{tabular}{c|c|c|c|c|c|c}
    \hline
        z & $E_{B-V}$ & $E_{B-V}$ & $E_{B-V}$ & completeness & purity\\
         input & input & prior & upper limit &  & \\
        \hline
         \emph{uniform} & \emph{none} & \emph{none} & \emph{none} & \emph{\textbf{98.1\%}} & \emph{99.4\%}\\
         \emph{uniform} & \emph{none} & \emph{basic} & \emph{none} & \emph{\textbf{85.4\%}} & \emph{99.7\%}\\
         uniform & basic & basic & none& \textbf{84.3\%} & 92.0\%\\
         uniform & evolving & evolving & none & \textbf{81.7\%} & 99.8\%\\
         uniform & evolving & evolving & upper limit 1 & \textbf{88.0\%} & 98.0\%\\ 
         uniform & evolving & evolving & upper limit 2 & \textbf{84.3\%} & 84.3\%\\
         \hline
         \emph{expected} & \emph{none} & \emph{none} & \emph{none} & \emph{97.3\%} & \emph{\textbf{99.7\%}}\\
         \emph{expected} & \emph{none} & \emph{basic} & \emph{none} & \emph{88.9\%} & \emph{\textbf{> 99.99\%}}\\ 
         expected & basic & basic & none & 78.9\% & \textbf{94.0\%}\\
         expected & evolving & evolving & none & 87.9\% & \textbf{99.4\%}\\
         expected & evolving & evolving & upper limit 1 & 77.3\% & \textbf{96.1\%}\\
         expected & evolving & evolving & upper limit 2 & 82.4\% & \textbf{83.8\%}\\
         \hline
    \end{tabular}
    \label{tab:allResults}
\end{table*}

\subsection{Completeness}

Completeness is a key metric for estimating the instrument performance at high redshifts by determining how many high-z GRBs will be correctly identified as such and how many will be missed. In Table~\ref{tab:allResults}, we show the simulation results for both a uniform and an expected redshift distribution, but to estimate completeness we use the uniform distribution. 

\begin{figure}
    \centering
    \includegraphics[width = \linewidth]{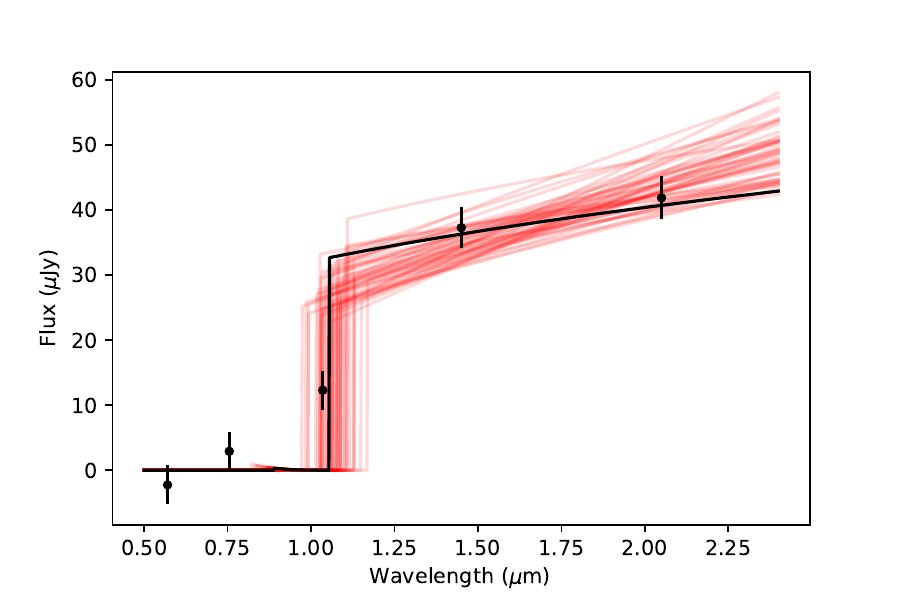}    \includegraphics[width = \linewidth]{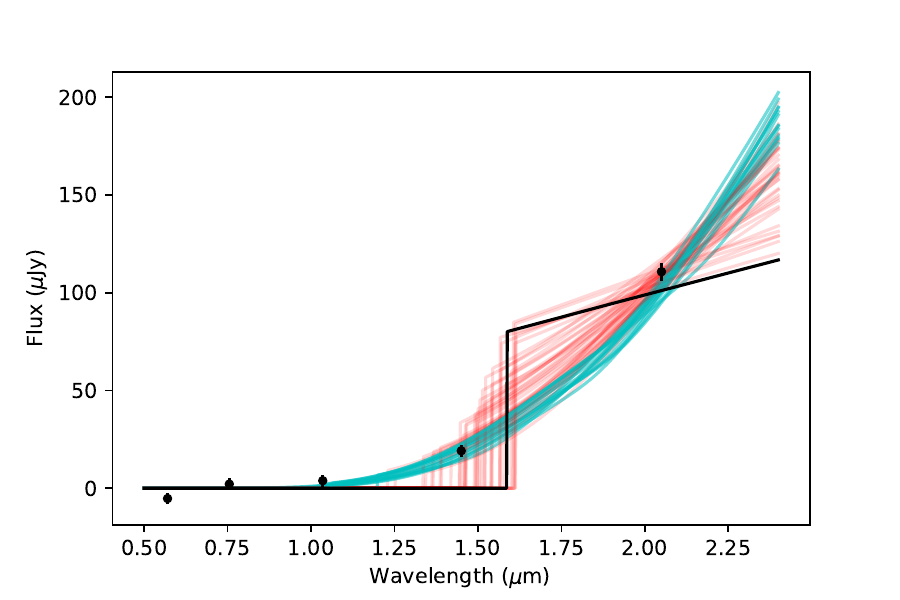}
    \caption{Example fits to simulated spectra, with the black lines showing the input GRB spectra, the black points showing simulated fluxes with 1$\sigma$ uncertainties, and the blue lines indicating the final positions of the 50 walkers. High-redshift solutions are in red, and low-redshift solutions are in blue. \textbf{Top:} A fit where all 50 walkers retrieve a redshift within 10\% of the input redshift. \textbf{Bottom:} A fit where the walkers find varying solutions. Both the high-z and low-z high-extinction outputs adequately fit the data, giving a mixed result for the GRB redshift.
    }
    \label{fig:confusion}
\end{figure}

For the `no extinction baseline' run, the completeness is unsurprisingly high at 98.1\% since the high-z GRBs cannot be mistaken for high-extinction GRBs in this case. 
Of the $\sim 2\%$ of GRBs that are mistaken as low-redshift, all had redshifts between $z =$ 5 - 5.5, and were mistaken as GRBs with redshifts between $z = $ 4.5 - 5, so they were only missed due to their proximity to the high-redshift threshold. 
When performing the `extinction prior baseline' run, the completeness drops significantly to 85.4\%.
Including extinction in the fitting method increases the chances of confusing high-z GRBs with low-z dusty interlopers, especially when there are only one or two filters with non-zero fluxes (see Figure \ref{fig:confusion}). 

When including extinction as both an input and fitting parameter, the `evolving' extinction distribution with upper limit 1 (see Table~\ref{tab:EbvPriors}) performed the best with a completeness of $88.0\%$.
However, for all simulations that use extinction for both the inputs and fit priors, the completeness is relatively similar, between $\sim 82-88\%$, so the choice of distribution seems to have a limited effect on the completeness.

Only taking GRBs with $5 < z < 13$ has little impact on the completeness result. The method does seem to be prone to missing high-z GRBs when using the `evolving' extinction prior.
For example, when using the `evolving' prior without upper limits, GRBs with $z > 13$ account for 43\% of all missed high-z GRBs, while $z > 8.9$ account for 80\% of all misidentified high-z GRBs (See Figure \ref{fig:evolvingExtinction}:Left).
GRBs with $z > 8.9$ have their Ly$\alpha$ dropoff occur in one of the two reddest photometric bands, where it is more difficult to distinguish between high-z and high-extinction GRBs (see Figure \ref{fig:confusion}). 

When examining the completeness between $5 < z < 13$, the completeness for the `evolving' extinction prior increases from 81.7\% to 83.0\%.
However, for GRBs with $5 < z < 8.9$, where most high-z GRBs are expected to be, the completeness increases to 93.0\%.
A large portion of the remaining miss-identified high-z GRBs are those near the high-z threshold.
Of the missed high-z GRBs with $5 < z < 13$, as many as $\sim10\%$ fall between 5 - 5.5 when using the `evolving' extinction with upper limit 1 (see Figure \ref{fig:special1evolvingExtinction}:Left).

\subsection{Purity}

The purity metric is used to determine the instrument's ability to correctly identify low-redshift GRBs.
It estimates the risk for false positives by mistaking low-redshift, high-extinction GRBs for high-redshift ones.
It is crucial to reduce the number of false positives, because if high-z detections are regularly false alarms, this will have a large impact on the use of follow-up resources. 
For estimating purity we focus on the expected redshift distribution because it evaluates the false positive rate for a general population of GRBs.

The baseline fits both have very high purity results of 99.7\% and $> 99.99\%$ for the `no extinction baseline' and `extinction prior baseline' simulations, respectively; so when extinction is not present in the GRB spectra, very few low-z GRBs are mistaken for high-z ones.
Additionally, for both of the baseline runs, all of the GRBs mistaken for high redshift had input redshifts between 4.5-5 and output redshifts between 5-5.5, so these lapses in purity are caused by GRBs with redshifts near our high-redshift threshold.

\begin{figure}
    \centering
    \includegraphics[width = \linewidth]{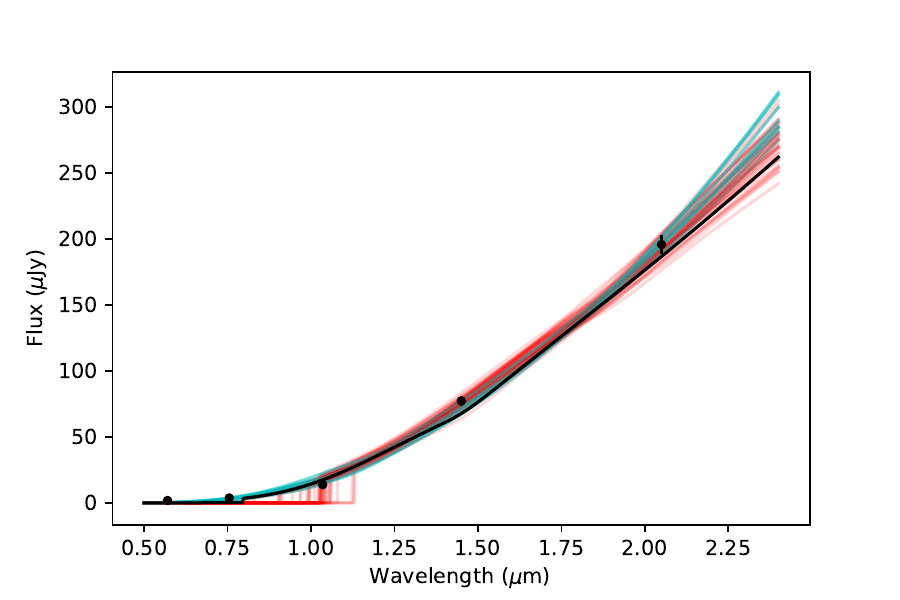}
    \includegraphics[width = \linewidth]{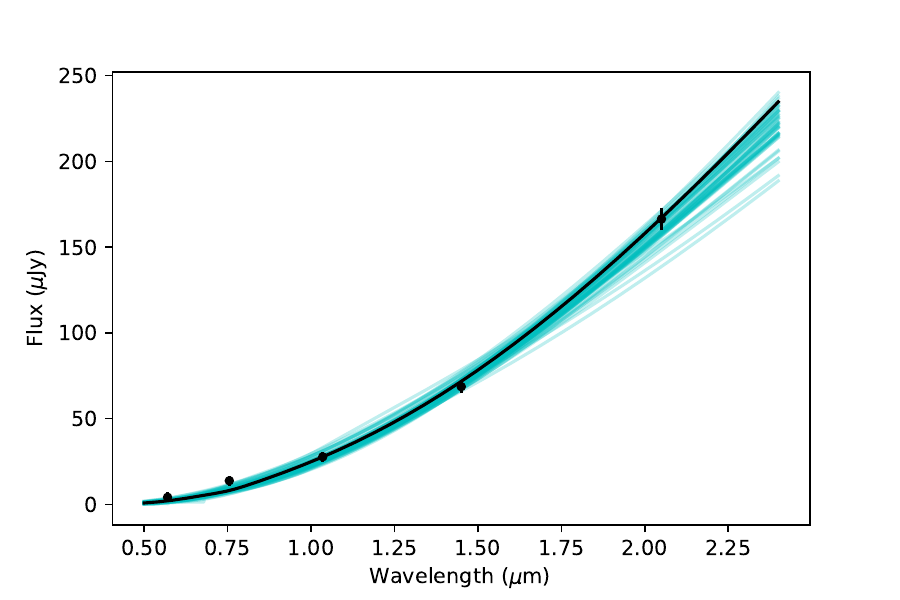}
    \caption{Example fits to simulated spectra, with the black lines showing the input GRB spectra, the black points showing simulated fluxes with 1$\sigma$ uncertainties, and the blue lines indicating the final positions of the 50 walkers. \textbf{Top:} An example of a fit using the `basic' extinction distribution for the inputs and priors. This simulated GRB has a high redshift and high extinction, which makes it difficult to determine the redshift accurately. High-redshift solutions are in red, and low-redshift solutions are in blue.
    \textbf{Bottom:} A similar fit using the `evolving' extinction distribution. The walkers are less inclined to combine high $z$ and high $E_{B-V}$ values for fitting, even though it would accurately fit the low flux in the bluer bands. This improves the redshift retrieval performance.}
    \label{fig:highzOopsie}
\end{figure}

When using the basic extinction distribution for both the inputs and the fitting, there is a drop in purity down to 94.0\%.
However, when switching to the `evolving' extinction prior, the purity jumps back up to 99.4\%, which is comparable to the baseline runs.
The improvement when switching from the `basic' to the `evolving' prior is likely due to the more realistic simulated GRB spectra and the additional information about extinction as a function of redshift.
When using the `basic' prior, there are more GRBs that have both a high redshift and a high extinction.
These kinds of simulated GRBs do not match up with observation based on studies done by \citet{Bolmer2018}, \citet{Covino2013}, and \citet{Greiner2011}, and have spectra that make it difficult to tease out the redshift.
The walkers are able to use both high redshift and high extinction for fitting because they do not have a prior that prevents them from settling on this unlikely solution.
Using the `evolving' extinction prior eliminates these kinds of GRBs and fits, which results in an increase in purity (see Figure~\ref{fig:highzOopsie}). 
We note that the implementation of upper limits on $E_{B-V}$ result in a decrease in purity, down to 96.1\% or even 83.8\% for the most constrained upper limit. This will be further discussed in Section~\ref{sec:paramSpace}.

\begin{table*}
\caption{Results from all simulations, listing redshift (z) and $E_{B-V}$ inputs and priors, as well as the accuracy from a Universe perspective ('Universe Acc.': for GRBs with an input redshift above 5, or between 5 and 12, how many have a retrieved redshift within 10\% or 20\% of the input redshift), and the accuracy from an observer perspective ('Observer Acc.': for GRBs with a retrieved redshift above 5, or between 5 and 12, how many are within 10\% or 20\% of the input redshift). Baseline runs are italicized.}
\begin{tabular}{c|c|c|c|c|c|c|c|c|c|c|c}
    \hline
        z & $E_{B-V}$ & $E_{B-V}$ & $E_{B-V}$ & \multicolumn{2}{c|}{Universe Acc. (10\%)} & \multicolumn{2}{c|}{Universe Acc. (20\%)} & \multicolumn{2}{c|}{observer Acc. (10\%)} & \multicolumn{2}{c|}{observer Acc. (20\%)}\\
         input & input & prior & upper limit & $5<z$ & $5<z<12$ & $5<z$ & $5<z<12$ & $5<z$ & $5<z<12$ & $5<z$ & $5<z<12$\\
        \hline
         \emph{uniform} & \emph{none} & \emph{none} & \emph{none} & \emph{78.2\%}  & \emph{93.5\%} & \emph{90.8\%} & \emph{96.8\%} & \emph{79.7\%} & \emph{95.5\%} & \emph{92.5\%} & \emph{99.1\%}\\
         \emph{uniform} & \emph{none} & \emph{basic} & \emph{none} & \emph{58.8\%} & \emph{70.3\%} & \emph{75.2\%} & \emph{81.6\%} & \emph{68.8\%} & \emph{75.5\%} & \emph{87.8\%} & \emph{87.8\%}\\
         uniform & basic & basic & none & 49.2\% & 50.7\% & 65.0\% & 62.7\% & 54.6\% & 56.6\% & 72.0\% & 70.6\%\\
         uniform & evolving & evolving & none & 62.2\% & 79.3\% & 75.0\% & 84.0\% & 76.0\% & 92.1\% & 91.7\% & 97.3\%\\
         uniform & evolving & evolving & upper limit 1 & 68.2\% & 85.4\% & 81.0\% & 89.4\% & 76.5\% & 91.3\% & 90.9\% & 95.8\%\\ 
         uniform & evolving & evolving & upper limit 2 & 64.5\% & 76.4\% & 77.6\% & 82.1\% & 70.3\% & 83.2\% & 84.7\% & 89.2\%\\
         \hline
         \emph{expected} & \emph{none} & \emph{none} & \emph{none} & \emph{90.4\%} & \emph{92.0\%} & \emph{94.5\%} & \emph{94.9\%} & \emph{93.0\%} & \emph{94.8\%} & \emph{97.1\%} & \emph{97.5\%}\\
         \emph{expected} & \emph{none} & \emph{basic} & \emph{none} & \emph{83.4\%} & \emph{83.8\%} & \emph{86.8\%} & \emph{87.1\%} & \emph{92.9\%} & \emph{93.7\%} & \emph{96.7\%} & \emph{97.2\%}\\ 
         expected & basic & basic & none & 62.8\% & 63.9\% & 71.6\% & 72.3\% & 54.9\% & 57.0\% & 63.2\% & 65.1\%\\
         expected & evolving & evolving & none & 83.5\% & 86.2\% & 87.0\% & 88.1\% & 91.9\% & 95.2\% & 95.6\% & 96.8\%\\
         expected & evolving & evolving & upper limit 1 & 73.3\% & 73.1\% & 76.9\% & 76.3\% & 74.9\% & 80.4\% & 79.0\% & 84.4\%\\
         expected & evolving & evolving & upper limit 2 & 78.4\% & 78.6\% & 82.5\% & 82.6\% & 47.1\% & 58.7\% & 49.7\% & 61.8\%\\
         \hline
    \end{tabular}
    \label{tab:accuracyResults}
\end{table*}

\subsection{Accuracy}   

The accuracy metric determines how well the code can estimate the actual redshift of a GRB. 
We explore the accuracy from two different perspectives. 
First we examine the accuracy from a `Universe perspective', i.e., for GRBs with input redshifts of $z>5$, how many have an accurate measured redshift. This shows how well the instrument can estimate the actual redshift.
Secondly, we examine the accuracy from an `observer perspective', i.e., for GRBs with measured redshifts of $z>5$, how many have accurately retrieved the input redshift. 
This tells us how many of the GRBs identified as having a high redshift will have accurate redshift measurements.

\begin{figure}
    \centering
    \includegraphics[width = \linewidth]{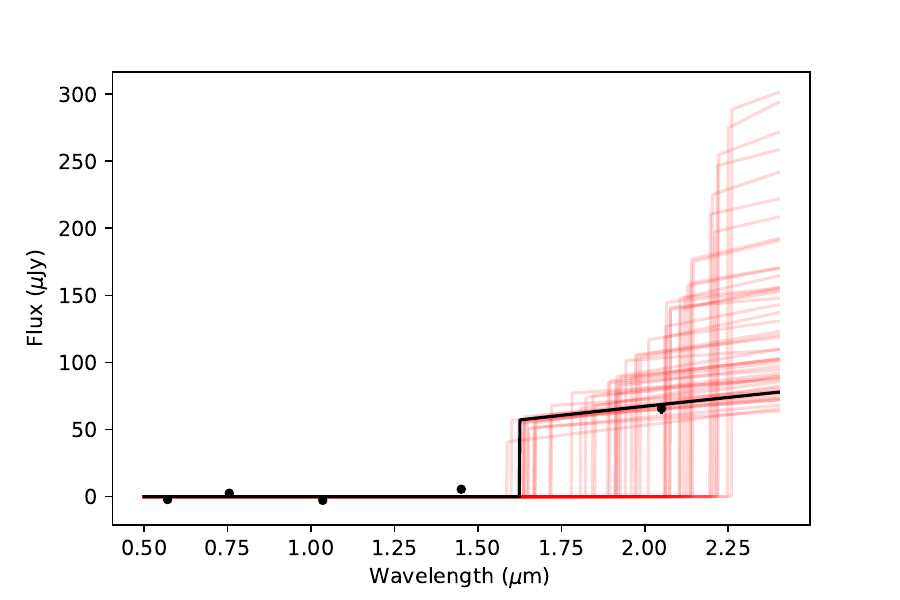}
    \caption{An example of a fit to a simulated $z\sim 12.4$ GRB from the `no extinction baseline' run, with the black lines showing the input GRB spectra, the black points showing simulated fluxes with 1$\sigma$ uncertainties, and the blue lines indicating the final positions of the 50 walkers. For GRBs where the Ly$\alpha$ dropoff occurs in or near the reddest wavelength band (z > 12), it becomes very difficult to constrain the redshift.}
    \label{fig:highzfit}
\end{figure}

For the `no extinction baseline', when looking at the accuracy from the Universe perspective and GRBs with redshifts above 5, we retrieve a redshift within 10\% of the input redshift in 78.2\% of the cases for a uniform redshift distribution, and in 90.4\% of the cases for an expected redshift distribution.
The large accuracy difference between the uniform and expected redshift distributions can be attributed to the much lower abundance of $z > 12$ GRBs in the expected redshift distribution.
The method struggles at redshifts above $z\sim 12$ because either only the reddest filter has a non-zero flux (for $z > 13$) or the flux in the second reddest band is low enough that it is easily mistaken as a non-zero flux measurement (for $12<z<13$).
In these cases, constraining the fit parameters becomes very difficult (see Figure~\ref{fig:highzfit}).
When looking only at the redshift retrieval between $z$ of $5-12$ for the uniform distribution, the accuracy increases to 93.5\%, which is comparable to the result for the expected distribution.
From an observer perspective, the accuracy is comparable to the accuracy from a Universe perspective for all accuracy metrics in the case of the `no extinction baseline' (see Table \ref{tab:accuracyResults}).

When using the `extinction prior baseline', the accuracy from the Universe perspective for the uniform redshift distribution drops to 58.8\%. This sharp loss of accuracy can be attributed to the increase in the complexity of the model.
When adding extinction as a free parameter, at least 3 filters with non-zero flux measurements are required to constrain the flux, redshift and extinction.
The fitting method has a much harder time constraining the redshift for GRBs with $z > 8.9$ where the Ly$\alpha$ dropoff falls in one of the two reddest filters and only 1-2 non-zero fluxes are present (see Figure~\ref{fig:noInputbasicFit}:Left).
For the expected distribution, there is also a drop in accuracy, but it is not nearly as severe as the loss of accuracy for the uniform redshift distribution: from 90.4\% to 83.4\%.
This is because the expected redshift distribution has far fewer $z > 8.9$ GRBs than the uniform distribution, and therefore is less impacted by the addition of extinction as a fitting parameter.
The accuracy from an observer perspective also decreases, but not nearly as much as the accuracy from a Universe perspective.
For example, for GRBs with measured redshifts of $z>5$, the code retrieves the redshift within 10\% of the input redshift 68.8\% of the time, compared to 58.8\% of the time from a Universe perspective.

When extinction is included as both an input and fitting parameter using the `basic' extinction distribution, the accuracy continues to decrease from both a Universe and observer perspective. This is due to high-extinction GRB spectra now present in the sample, which have low relative flux values in the bluer bands and are thus easy to mistake for a high-redshift GRB. 
When using the `evolving' extinction distribution, the instrument's accuracies become comparable to those of the `extinction prior baseline'. 
Of the different `evolving' extinction distributions, the performance of the one with no upper limits on $E_{B-V}$, and the one using upper limit 1 for $E_{B-V}$ (see Table \ref{tab:EbvPriors}), are comparable to that of the `extinction prior baseline' simulation for nearly every accuracy measurement, regardless of input redshift distribution.
This is promising because the instrument is expected to perform as well as it would in the best case scenario where every detected GRB has virtually no extinction.

\begin{figure}
    \centering
    \includegraphics[width = \linewidth]{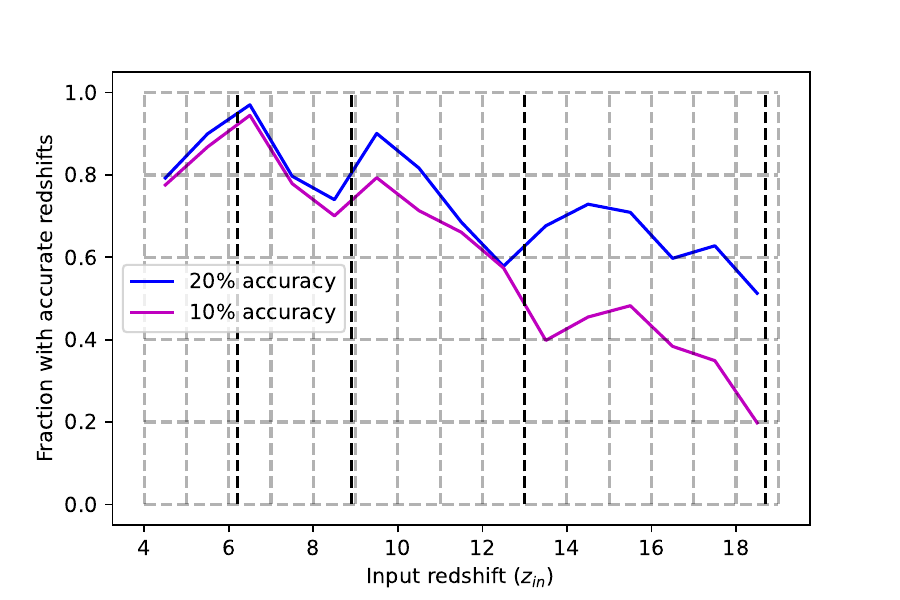}
    \includegraphics[width = \linewidth]{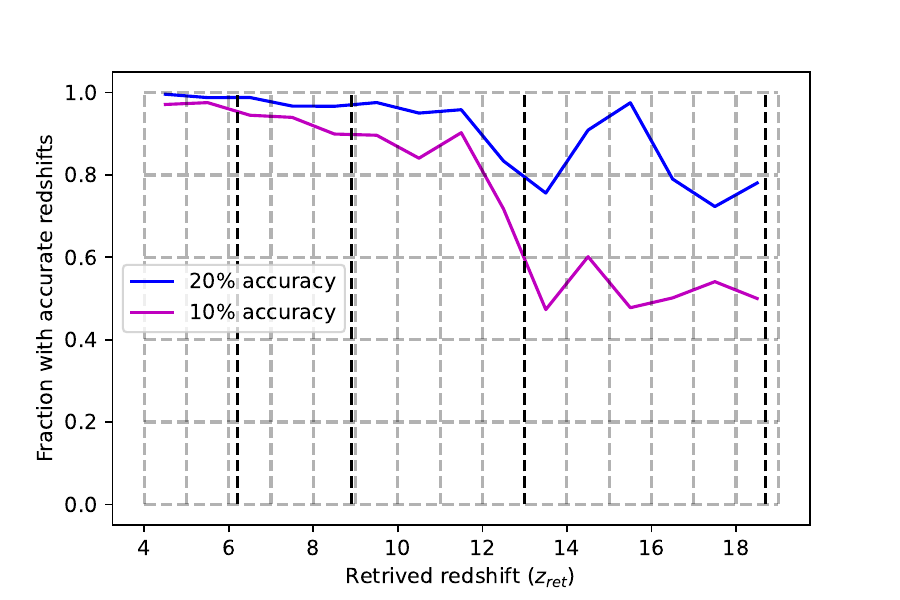}
    \caption{Accuracy as a function of redshift when using the `evolving' extinction prior without upper limits. The black dashed lines represent the redshifts where the Ly$\alpha$ dropoff aligns with the filter edges, and the grey dashed lines show redshift in increments of 1 and the fraction of GRBs with accurate redshifts in increments of 0.2. \textbf{Top:} The fraction of retrieved redshifts within 10\% and 20\% of the input redshift for $z_{in} > 5$ (Universe perspective). \textbf{Bottom:} The fraction of retrieved redshifts within 10\% and 20\% of the input redshift for $z_{ret} > 5$ (observer perspective).}
    \label{fig:evolving0accuracy}
\end{figure}

\begin{figure}
    \centering
    \includegraphics[width = \linewidth]{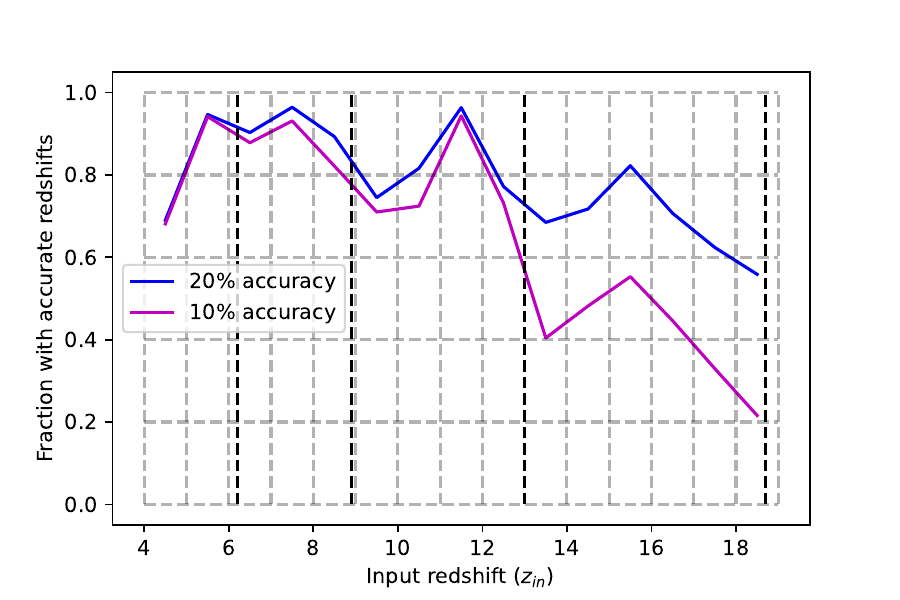}
    \includegraphics[width = \linewidth]{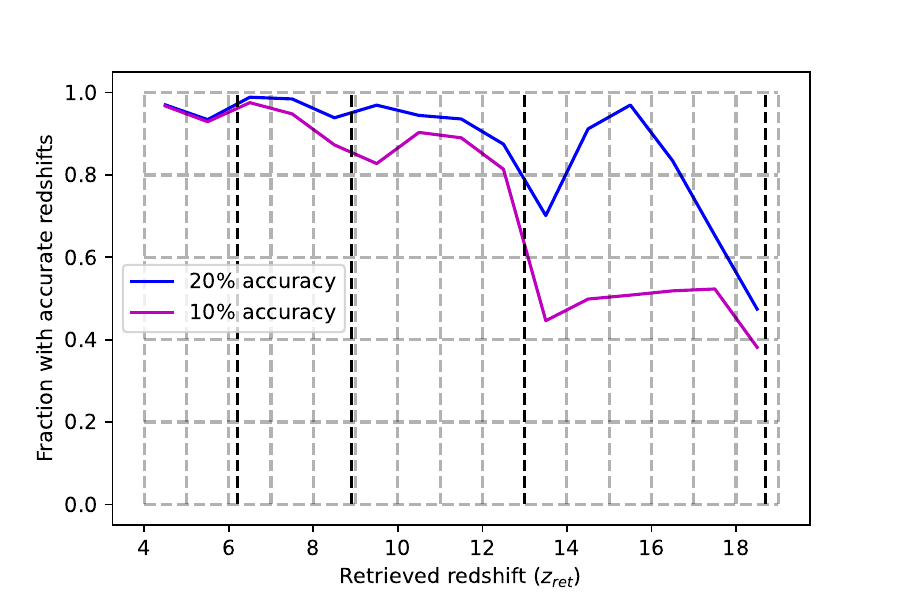}
    \caption{Accuracy as a function of redshift when using the `evolving' extinction prior with upper limit 1. The black dashed lines represent the redshifts where the Ly$\alpha$ dropoff aligns with the filter edges, and the grey dashed lines show redshift in increments of 1 and the fraction of GRBs with accurate redshifts in increments of 0.2.  \textbf{Top:} The fraction of retrieved redshifts within 10\% and 20\% of the input redshift for $z_{in} > 5$ (Universe perspective). \textbf{Bottom:} The fraction of retrieved redshifts within 10\% and 20\% of the input redshift for $z_{ret} > 5$ (observer perspective).}
    \label{fig:evolving1accuracy}
\end{figure}

The run using the `evolving' extinction prior with upper limit 1 tends to do slightly better from the Universe perspective, as for GRBs with $5 < z < 12$ it retrieves a redshift within 10\% of the input redshift in 85.4\% of cases, compared to 79.3\% of cases for the `evolving' extinction prior without upper limits. 
However, using the `evolving' extinction prior with no upper limits outperforms all other runs from an observational perspective.
For example, for GRBs with a measured redshift $5 < z < 12$, the measured redshift is within 10\% of the input redshift 92.1\% of the time when using the `evolving' extinction without upper limit, as opposed to 91.3\% of the time when using the `evolving' extinction with upper limit 1.
This pattern holds when examining the accuracy metric as a function of redshift (see Figures \ref{fig:evolving0accuracy} and \ref{fig:evolving1accuracy}).
For the `evolving' extinction prior without upper limits, the accuracy from a Universe perspective gradually decreases with higher redshifts, but from an observer perspective it has a high accuracy out to $z\sim 12$.
Conversely, the `evolving' extinction prior with upper limit 1 has a more reliable accuracy at higher redshifts from a universe perspective, but has a less reliable accuracy from an observational perspective.
This means that while the `evolving' extinction with upper limit 1 has a better accuracy for high-z GRBs overall, for GRBs that are successfully identified as high-redshift the `evolving' extinction prior without upper limits is more reliable.
For full accuracy results see Table \ref{tab:accuracyResults}.

\section{Discussion}
\label{sec:discussion}

In the previous section, we presented a variety of statistics for completeness, purity, and accuracy when making different assumptions about the redshift and host galaxy extinction distributions. 
We will now discuss the implications of these results for determining which extinction prior leads to the best performance of the \emph{Gamow} PIRT, the drawbacks of implementing $E_{B-V}$ upper limits, the limitations of the instrument at low redshifts, and the advantages of simultaneous observations in different photometric bands.

\subsection{Optimal $E_{B-V}$ prior}
\label{sec:optimalPrior}
There is a natural trade-off between completeness and purity. We have shown that adding upper limits on the host galaxy extinction increases the completeness while at the same time negatively affecting the purity. When looking at percentage changes, it is not obvious what the optimal choice of extinction model is. However, it is important to note that there is a much larger population of low-z GRBs, so a small change in percentage can have a noticeable impact on the number of false positives.
We must be mindful of this when balancing our ability to maximize the number of correctly identified high-z GRBs while minimizing false positives.
Here we use \emph{Swift}'s detection rate to estimate the number of high-z triggers and false alarms.

Since its launch in 2004, \emph{Swift} has detected $ \sim 100$ GRBs per year \citep{Gehrels2007}.
If we assume the same rate of detection for a mission like \emph{Gamow}, based on the measured redshift distribution \citep{Ghirlanda2015,Ghirlanda2021,Ghirlanda2022}, approximately 7.3 of these GRBs would have $z>5$, while 92.7 GRBs would have $z <5$. 
When using the `evolving' extinction prior with no upper limit, we find a completeness of 81.7\% and a purity of 99.4\%.
This translates to $\sim 6$ correctly identified high-z GRBs per year, and $\sim 0.2$ misidentified low-redshift GRB per year, or $\sim 1$ over the course of a 5-year mission.

Conversely, when using the `evolving' extinction prior with upper limit 1, we find a completeness of 88.0\% and a purity of 96.1\%, which translates to $\sim 6.4$ correctly identified high-z GRBs and $\sim 3.6$ misidentified low-z GRBs per year.
The small decrease in purity when using upper limit 1 results in a detrimental increase in false alarms, as now 36\% of reported high redshift GRBs would turn out to be low-z dusty interlopers.
For this reason, we would recommend to use a `evolving' extinction prior without upper limits for both simulating performance beforehand and fitting photometric data during the instrument's operation.

\subsection{Extremes of parameter space}
\label{sec:paramSpace}
For all simulations there is a loss of accurate redshift retrieval below $z\sim3$.
This result is expected as the Ly$\alpha$ break does not fall in any of the \emph{Gamow} PIRT photometric bands for GRBs with $z \leq 3.1$, so there are no features to distinguish between GRBs with redshifts below this threshold.
We are not concerned about the loss of accuracy for low-z GRBs as long as we are capable of accurately identifying them as having a low redshift.
The main goal for a high-z GRB mission is to minimize the rate of false positives caused by low-z dusty interlopers, so maximizing the purity is far more important than accurate redshift retrieval for $z < 3$ GRBs.

For the `evolving' extinction prior, we find that the more restrictive the upper limits on $E_{B-V}$, the worse the purity becomes (See Table \ref{tab:allResults}).
This degradation of purity is also visible in the input versus output redshift density plots.
See Figures~\ref{fig:evolvingExtinction}:Right, \ref{fig:special1evolvingExtinction}:Right and \ref{fig:special2evolvingExtinction}:Right for comparison of results from the `evolving' extinction runs with no upper limit, upper limit 1, and upper limit 2, respectively. 
This occurs because the upper limits restrict the walkers to regions of parameter space that they would otherwise have access to at a different redshift.
For walkers with a high-redshift guess at the start of the burn-in phase, they are restricted to fitting the fluxes with a lower $E_{B-V}$, which may inadvertently push them towards a higher redshift if the bluer bands have relatively low flux values.
While this is not an issue when fitting high-z GRBs, it is problematic for walkers fitting a low-z high-extinction GRB if they start in this region of parameter space.
These upper limits are an example of a prior that, while adding more information about the parameters, can impact the walkers' ability to move through parameter space.

\subsection{Performance for low flux GRBs}

\begin{figure}
    \centering
    \includegraphics[width = \linewidth]{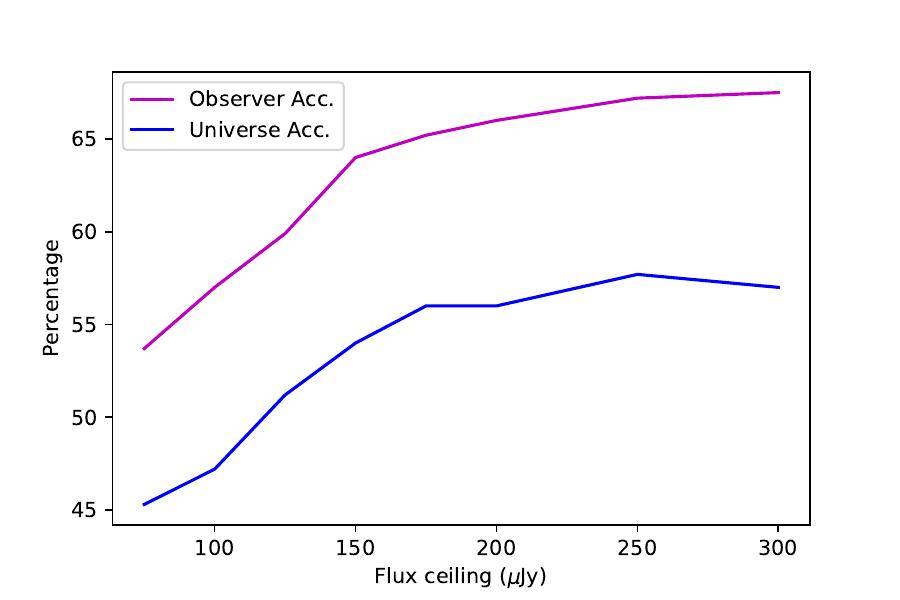}
    \caption{Accuracy as a function of flux for both a Universe and an observer perspective. The accuracy is the rate of retrieval within 10\% of the relevant redshift. The flux ceiling is the maximum flux in the reddest photometric band of the GRBs considered.}
    \label{fig:lowFluxAcc}
\end{figure}

When examining the completeness, purity, and accuracy statistics for GRBs with lower flux values, the purity and accuracy tend to decrease, while the completeness remains relatively stable. 
For example, when using the 'evolving' extinction prior with no upper limits, the percentage of retrieved redshifts within 10\% of the input redshift for GRBs with $z > 5$ drops to 45.3\% when only looking at those with fluxes less than 75~$\mu$Jy in the reddest photometric band, compared to 79.3\% for all GRBs with $z > 5$.
The accuracy also decreases at low flux from an observer perspective.
When looking at GRBs with $z > 5$ and fluxes in the reddest band less than 75 $\mu$Jy, only 53.7\% of retrieved redshifts are within 10\% of the input redshift for GRBs, compared to 91.9\% for all GRB with $z > 5$ (see Figure \ref{fig:lowFluxAcc}).

\begin{figure}
    \centering
    \includegraphics[width=\linewidth]{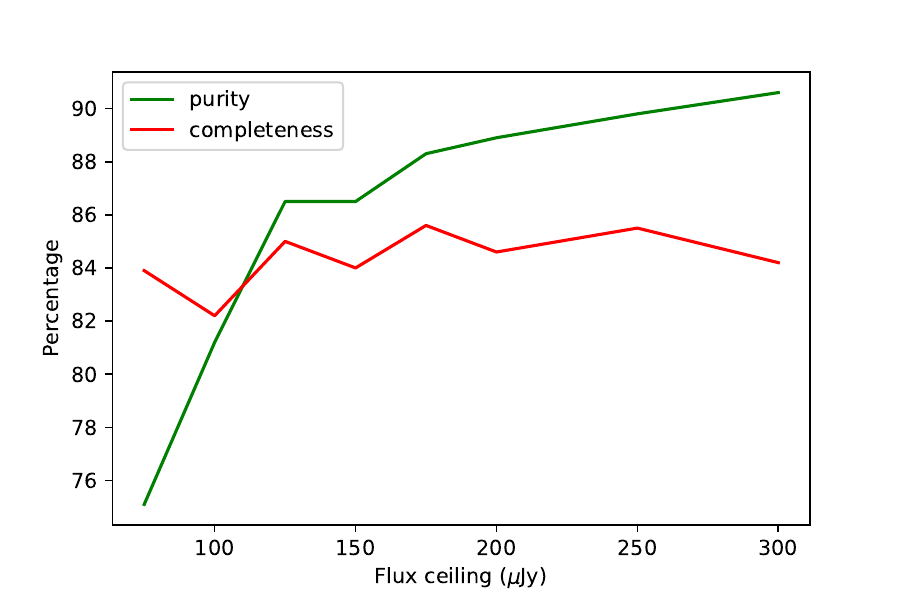}
    \caption{Completeness and purity as a function of flux. The flux ceiling is the maximum flux in the reddest photometric band of the GRBs considered. The purity decreases at lower flux values, while the completeness is fairly constant.}
    \label{fig:lowFluxPC}
\end{figure}

The purity also drops significantly from 99.4\% for all GRBs to 81.2\% for GRBs with fluxes less than 100 $\mu$Jy, and to 75.1\% for GRBs with fluxes less than 75~$\mu$Jy in the reddest band.
However, the completeness remains largely unaffected regardless of GRB flux, as it hovers between 82-88\% when looking at GRBs with fluxes less than 75~$\mu$Jy up to GRBs with fluxes less than 300~$\mu$Jy in the reddest band (see Figure \ref{fig:lowFluxPC}).
This is also consistent with the behavior of the completeness when using different extinction priors.

While the overall purity is very high and the rate of false positives is low, GRBs with lower flux values are more likely to be confused as having a high redshift. 
For this reason, it may be important to flag low-flux bursts as being higher risk targets for follow-up observations of future high-z GRB missions.

\subsection{Simultaneous bands}

In recent years, multiple instruments have been designed to use dichroics and beam splitters to take simultaneous photometric band measurements, such as GROND \citep{Greiner2008}, the PIRT on the \emph{Gamow} explorer \citep{Seiffert2021, White2021}, and SCORPIO, an imaging and spectroscopy instrument in development for the Gemini South telescope \citep{Robberto2020}.
Simultaneous measurements in different photometric bands are advantageous for fast evolving transients such as GRBs, because it eliminates the uncertainty that arises when correcting for rapid fading of the afterglow: even a small change in time can result in a large change in flux, which can only be corrected for by interpolation and extrapolation.
Even careful correction can introduce error in the photometric band fluxes, as it still relies on accurate measurements of temporal indices to do the correction. This is particularly important for the fast-evolving light curves of GRBs. 
There is also no guarantee that the light curve will have a steady decay between observations, due to for instance flares \citep{Greiner2009,Gao2009} or rebrightening episodes \citep{Kann2018,Bersier2003}. 
It is important to minimize any extra sources of error or uncertainty for high-z GRB missions where accurate photometric estimation is crucial, and any offsets or uncertainty in the bands relative fluxes could have an impact on redshift estimation. 
Even a small change in the purity can have a drastic impact on false alarm rates for a high-z GRB mission (see Section~\ref{sec:optimalPrior}), so it is imperative to reduce the likelihood of confusion by eliminating as many sources of error as possible. 
The issue of interpolation could be partly mitigated with dense temporal sampling, but simultaneous photometric bands will be the optimal solution for future high-redshift GRB missions. 

\section{Conclusions}
\label{sec:conclusion}

GRBs are valuable probes of the high-redshift Universe due to their high luminosities and simple power-law spectra. 
They are ideal for tracking the chemical evolution of the Universe, studying early star formation, and constraining the end of the Epoch of Reionization.
Multiple missions including optical-NIR photometric instruments have been proposed for finding and studying high-redshift GRBs.
Host galaxy extinction poses a challenge for constraining GRB redshifts, as low-z high-extinction GRBs can mimic high-z GRBs when using broad photometric bands.
It is imperative that future high-z GRB missions are capable of rapidly and reliably identifying the redshift of detected GRBs with as few false alarms as possible to encourage community follow-up.

\texttt{phozzy} is a photo-z simulations and fitting code that can be used by future high-z GRB missions for testing their ability to accurately retrieve the redshift of an expected GRB population.
Using it, we have tested the capabilities of the \emph{Gamow} PIRT when using different extinction priors, and found that using the `evolving' $E_{B-V}$ distribution gave the best results.
We find that \emph{Gamow} PIRT would have a $\sim92\%$ accuracy from an observer perspective, a completeness of $\sim82\%$, and would only mistake $\sim 0.6 \%$ of low-redshift GRBs as having a high ($z>5$) redshift.
This translates to $\sim 1$ false alarm per 500 GRBs detected.
We have also shown that we can increase the completeness to $\sim88\%$ by imposing constraints on the $E_{B-V}$ at high redshifts, but this would have a significantly negative effect on the purity. The latter would decrease such that $\sim1/3$ of high-redshift GRB alerts would in fact be a low-z dusty interloper. 
We note that the completeness for the `evolving' $E_{B-V}$ distribution increases to well above $90\%$ for GRBs with $5<z<9$. 
Finally, we discuss that the use of simultaneous measurements in different photometric bands can help remove unnecessary sources of error and improve the chances of retrieving GRB redshifts.

\section*{Acknowledgements}

We would like to acknowledge Alex Kann for his contributions to the work presented here and the development of the Gamow mission, and his remarkable contributions to GRB science in general.

We would like to thank the anonymous referee for their constructive feedback.
We would also like to thank the George Washington University's Office of the Vice Provost for Research and Dean's Office of the Columbian College of Arts \& Sciences for their support of this project and the Gamow mission proposal preparation.
The research was carried out in part at the Jet Propulsion Laboratory, California Institute of Technology, under a contract with the National Aeronautics and Space Administration (80NM0018D0004).

\section*{Data Availability}

The data presented in this paper were generated with the publicly available \texttt{phozzy} code which is available for download at https://github.com/hmfausey/phozzy.git.
 



\bibliographystyle{mnras}
\bibliography{Bib} 




\appendix

\section{Redshift Retrieval Plots}

Here we present the 2D histograms for the input versus output redshifts found by \texttt{phozzy} for each simulation shown in Table~\ref{tab:allResults}, with input redshift on the x-axis, output redshift on the y-axis, and the colors corresponding to the number density in each bin.
Each figure corresponds to a different configuration of extinction inputs and priors.
In each figure, the uniform redshift distribution results are displayed on the left while the expected redshift distribution results are on the right.
For each of the plots, the solid white lines correspond to $z=5$, which we use as the boundary between low- and high- redshift GRBs.
The vertical dashed lines represent the redshifts where the Ly$\alpha$ line occurs at a band edge (occurring at z = 3.1, 4.3, 6.2, 8.9, 13.0, and 18.7 from left to right).
The dash-dotted line is where the input redshift would exactly equal the output redshift, and the dotted lines show where the retrieved redshift is within 10\% of the input redshift.

\begin{figure*}
\textbf{Input: no extinction, Fitting: no extinction}

    \includegraphics[width = 0.49\linewidth]{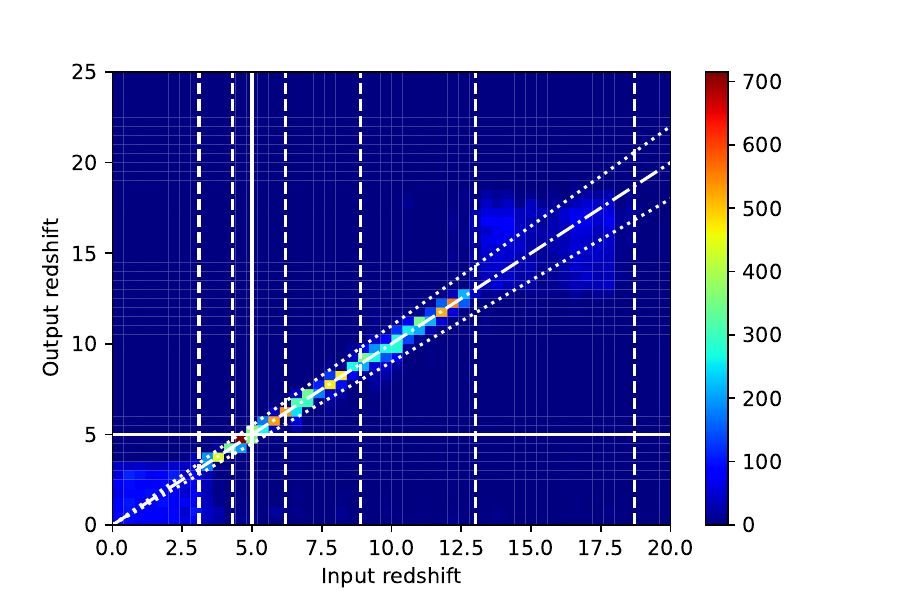}
    \includegraphics[width = 0.49\linewidth]{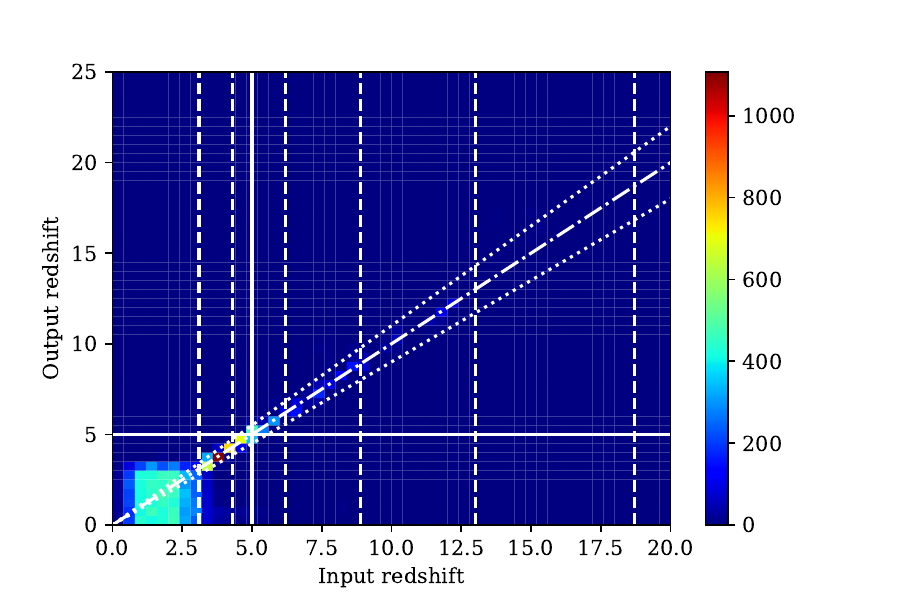}
    \caption{Input vs. output redshift when excluding extinction from both the inputs and priors. \textbf{Left:} uniform redshift distribution. \textbf{Right:} expected redshift distribution.
    }
    \label{fig:noE}    
\end{figure*}

\begin{figure*}
\textbf{Input: no extinction, Fitting: basic extinction}

    \centering
    \includegraphics[width = 0.49\linewidth]{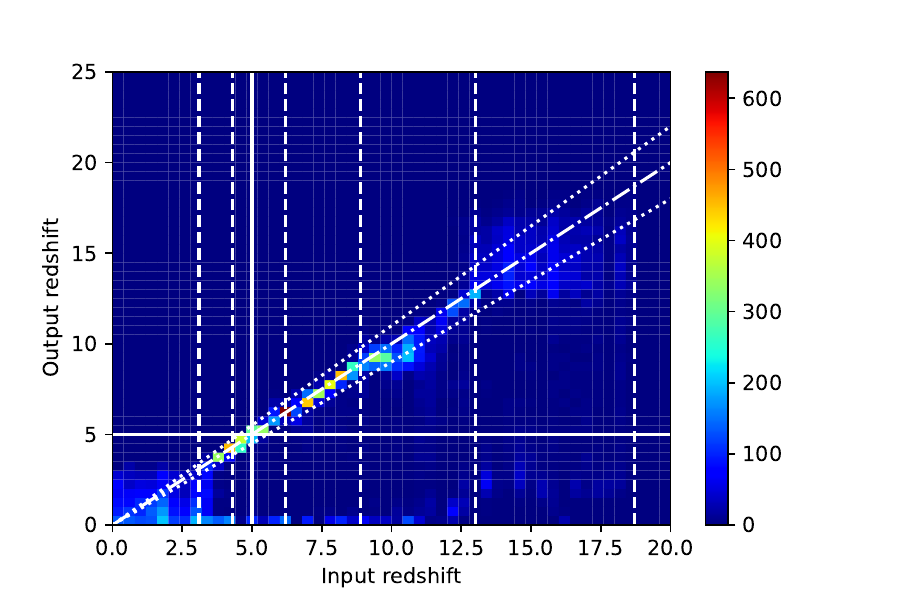}
    \includegraphics[width = 0.49\linewidth]{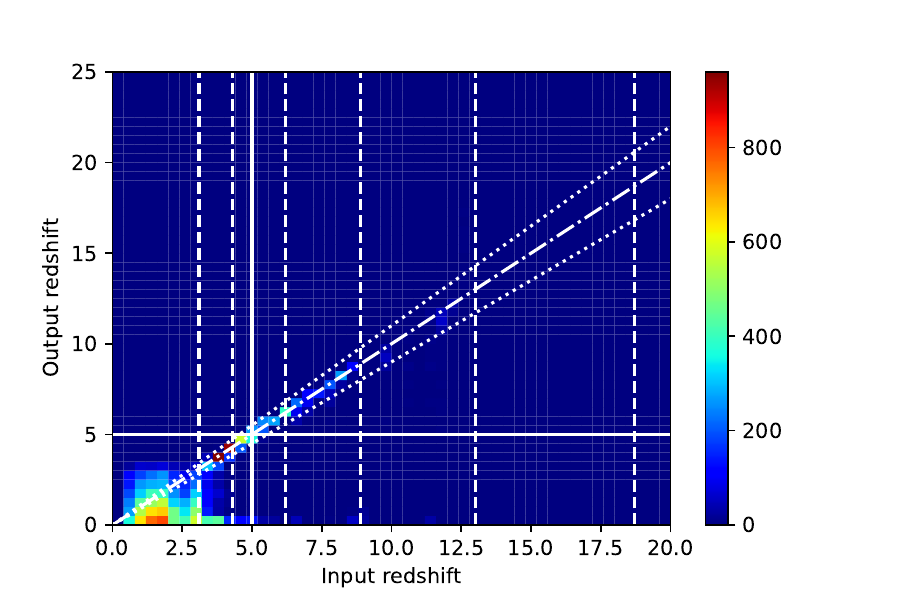}
    \caption{Input vs. output redshift when excluding extinction as an input parameter but including basic extinction prior for the fit. \textbf{Left:} uniform redshift distribution. \textbf{Right:} expected redshift distribution.
    }
\label{fig:noInputbasicFit}
\end{figure*}

\begin{figure*}
\textbf{Input: basic extinction, Fitting: basic extinction}

    \includegraphics[width = 0.49\linewidth]{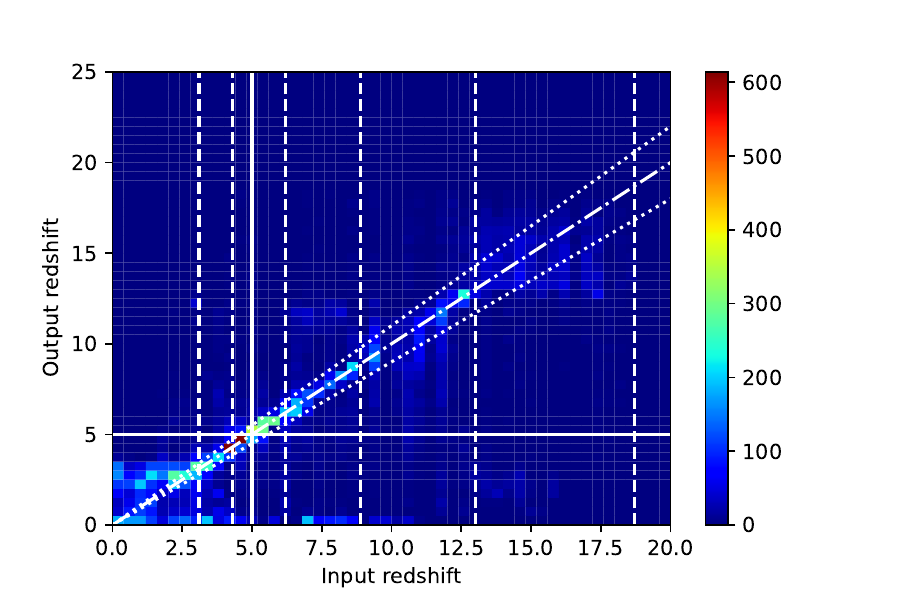}
    \includegraphics[width=0.49\linewidth]{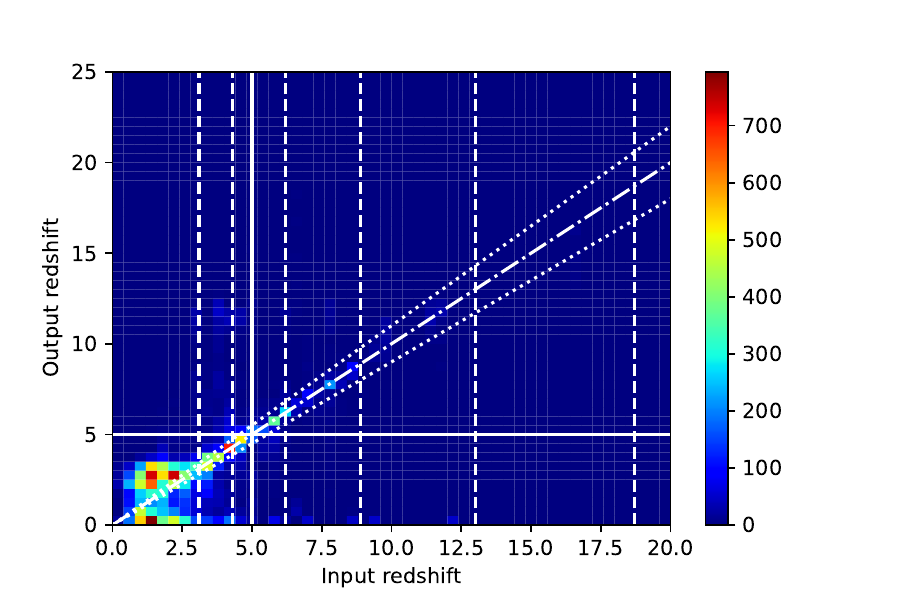}
    \caption{Input vs. output redshift when using a basic extinction distribution for both inputs and priors. \textbf{Left:} uniform redshift distribution. \textbf{Right:} expected redshift distribution.
    }
    \label{fig:basicInputandFit}
\end{figure*}

\begin{figure*}
\textbf{Input: `evolving' extinction, Fitting: `evolving' extinction}

    \includegraphics[width = 0.49\linewidth]{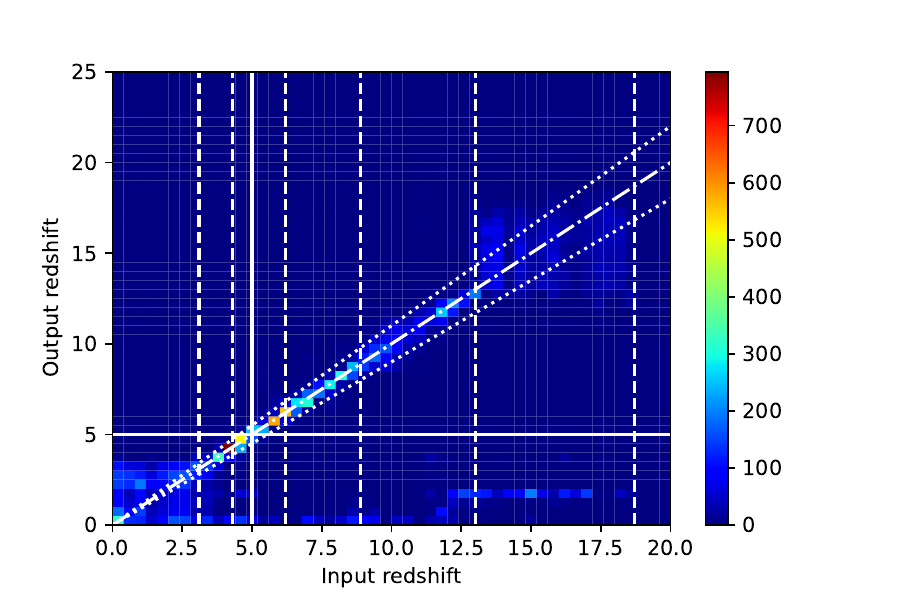}
    \includegraphics[width = 0.49\linewidth]{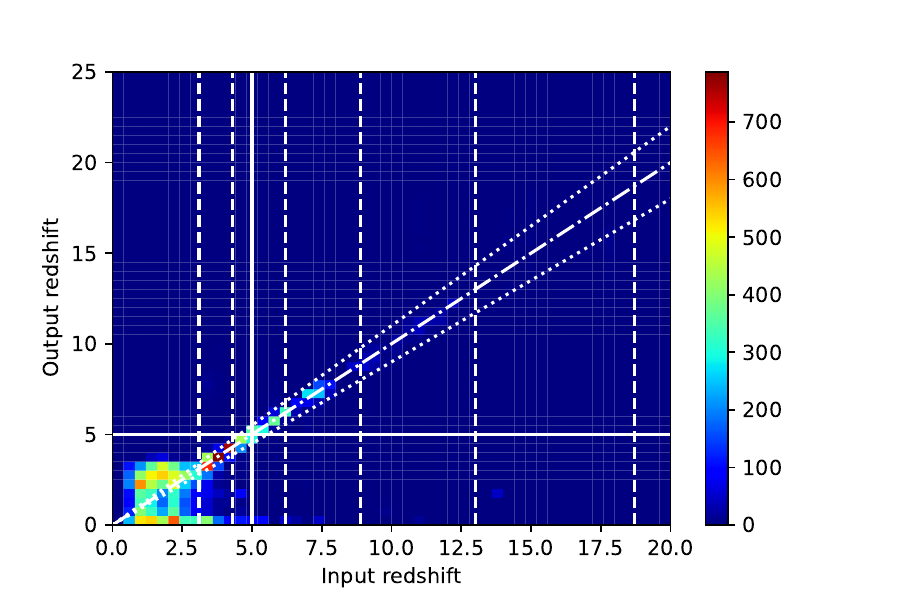}
    \caption{Input vs. output redshift when using the `evolving' extinction distribution for both input parameters and priors. \textbf{Left:} uniform redshift distribution. \textbf{Right:} expected redshift distribution.
    }
    \label{fig:evolvingExtinction}
\end{figure*}

\begin{figure*}
    \textbf{Input: `evolving' extinction with upper limit 1, Fitting: `evolving' extinction with upper limit 1}
    
    \includegraphics[width = 0.49\linewidth]{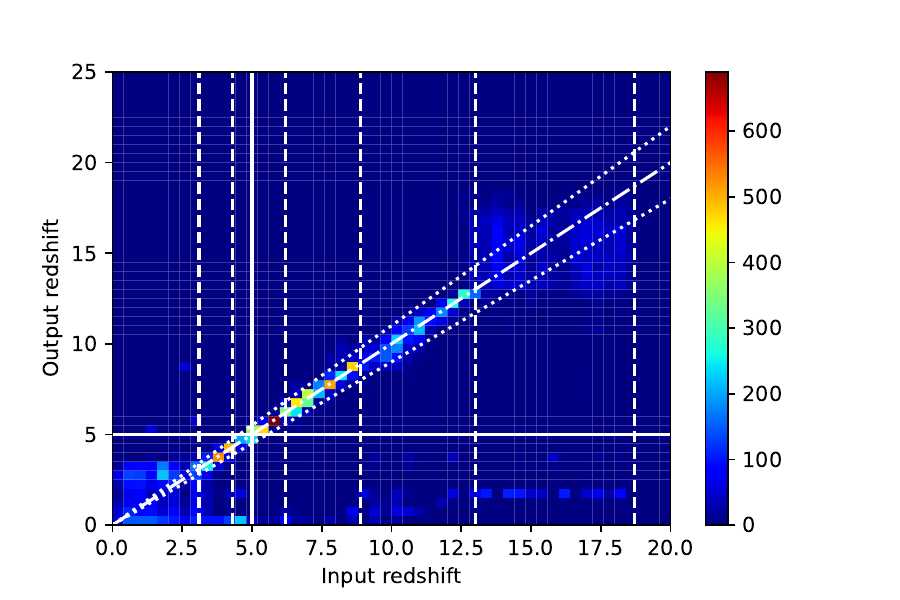}
    \includegraphics[width = 0.49\linewidth]{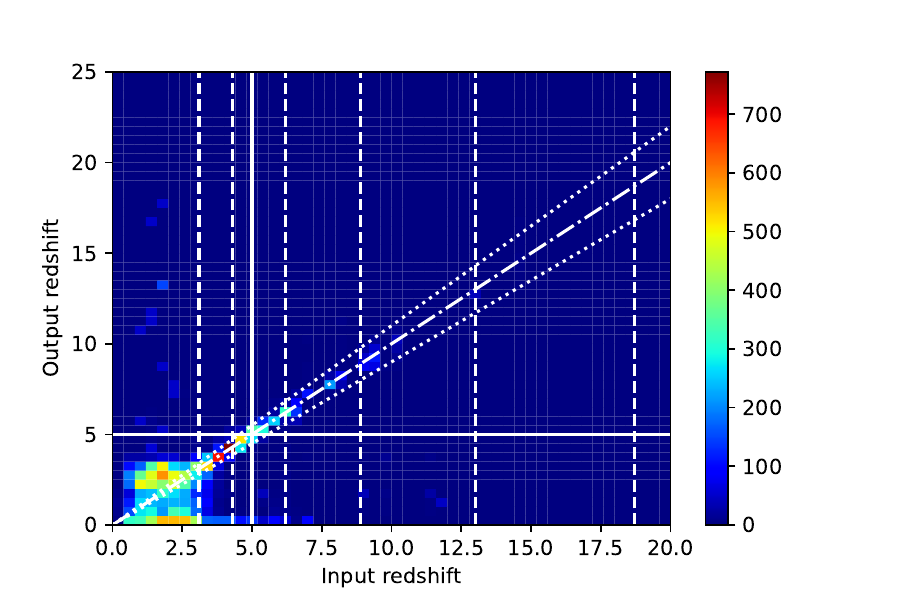}
    \caption{Input vs. output redshift when using the `evolving' extinction distribution for both input parameters and priors, and including upper limits for $E_{B-V}$ (see Figure \ref{tab:EbvPriors}:Upper limit 1). \textbf{Left:} uniform redshift distribution. \textbf{Right:} expected redshift distribution.
    }
    \label{fig:special1evolvingExtinction}
\end{figure*}

\begin{figure*}
    \textbf{Input: `evolving' extinction with upper limit 2, Fitting: `evolving' extinction with upper limit 2}
    
    \includegraphics[width = 0.49\linewidth]{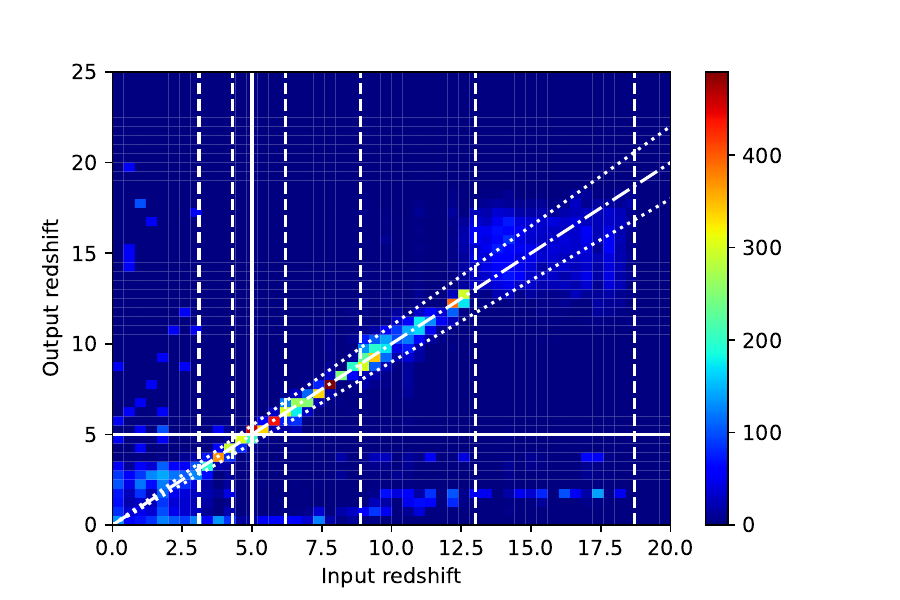}
    \includegraphics[width = 0.49\linewidth]{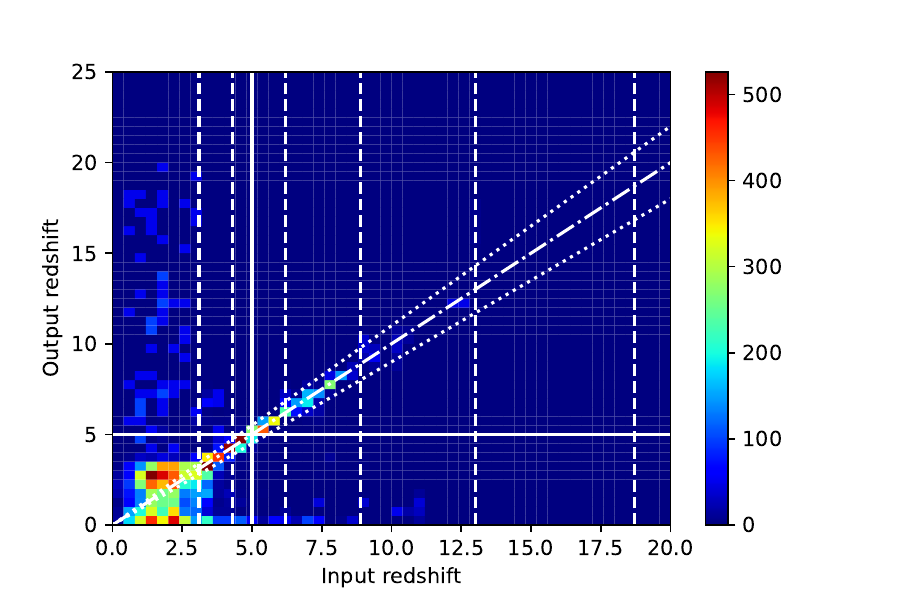}
    \caption{Input vs. output redshift when using the `evolving' extinction distribution for both input parameters and priors, including upper limits for $E_{B-V}$ (see Figure \ref{tab:EbvPriors}:Upper limit 2). \textbf{Left:} uniform redshift distribution. \textbf{Right:} expected redshift distribution.
    }
\label{fig:special2evolvingExtinction}
\end{figure*}

\bsp	
\label{lastpage}
\end{document}